\documentclass{aa}
\usepackage[breaklinks=true,colorlinks,citecolor=blue]{hyperref}
\usepackage{units}
\usepackage{txfonts}
\usepackage{graphicx}
\usepackage{threeparttable}
\usepackage{xcolor, tikz}
\usepackage[label font=bf,labelformat=simple]{subfig}
\usepackage{floatrow}
\bibpunct{(}{)}{;}{a}{}{,}

\newcommand{\modified}[1]{#1}  
\newcommand{\commented}[1]{}  

\newcommand{\feh}{\ensuremath{\rm [Fe/H]}\,}
\newcommand{\xfe}[1]{\ensuremath{\mathrm{[#1/Fe]}}\,}
\newcommand{\xh}[1]{\ensuremath{\mathrm{[#1/H]}}\,}

\definecolor{lime}{HTML}{A6CE39}
\DeclareRobustCommand{\orcidicon}{%
    \begin{tikzpicture}
    \draw[lime, fill=lime] (0,0) 
    circle [radius=0.16] 
    node[white] {{\fontfamily{qag}\selectfont \tiny ID}};
    \draw[white, fill=white] (-0.0625,0.095) 
    circle [radius=0.007];
    \end{tikzpicture}
    \hspace{-2mm}
}

\begin{document}

\title{TwinSpecNet: Extending APOGEE's chemical reach to low-S/N spectra via empirical paired learning}
\titlerunning{TSN: Paired learning for low-S/N APOGEE spectra}
\author{Weijia Sun\href{https://orcid.org/0000-0002-3279-0233}{\orcidicon}\inst{1}\thanks{E-mail: wsun@aip.de}, Cristina Chiappini\href{https://orcid.org/0000-0003-1269-7282}{\orcidicon}\inst{1,2} \and Samir Nepal\href{https://orcid.org/0000-0002-8557-5684}{\orcidicon}\inst{3,1}} 

\authorrunning{W. Sun \& C. Chiappini}

\institute{Leibniz-Institut für Astrophysik Potsdam (AIP), An der Sternwarte 16, 14482 Potsdam, Germany     \and{Laborat\'orio Interinstitucional de e-Astronomia - LIneA, Rua Gal. Jos\'e Cristino 77, Rio de Janeiro, RJ - 20921-400, Brazil} \and{Sydney Institute for Astronomy, School of Physics, University of Sydney, NSW 2006, Australia}}
        
\date{Received ; accepted }
\abstract
{Large spectroscopic surveys rely on automated pipelines to deliver homogeneous stellar labels, but a substantial fraction of observations are at low signal-to-noise ratio (S/N), where label estimates become imprecise or are omitted.
In APOGEE, these low-S/N spectra visits sample faint and distant populations---the bulge, outer halo, and satellite systems---yet still encode recoverable chemical information.
}
{We present TwinSpecNet (TSN), a paired-learning framework that exploits APOGEE's multi-visit observing strategy: by training on empirical low-/high-S/N spectral twins of the same stars, TSN learns to suppress stochastic noise while preserving the ASPCAP label scale.}
{TSN employs a Vision Transformer encoder with dual objectives: reconstructing high-S/N flux from low-S/N visits and predicting stellar parameters and abundances with calibrated uncertainties.
}
{TSN reduces label scatter relative to visit-level ASPCAP for $\mathrm{S/N}<60$ visits.
TSN reproduces the ASPCAP scale with residual scatters of $\sigma\simeq\unit[19]{K}$ in $T_{\mathrm{eff}}$, $\sigma\simeq\unit[0.06]{dex}$ in $\log g$, and $\sigma\simeq\unit[0.03]{dex}$ in \feh.
TSN tightens intra-cluster abundance dispersions, recovers cleaner chemical sequences in inner-disk and bulge and satellite samples, and improves C/N-based age precision for APOKASC giants from 1.70 to \unit[1.49]{Gyr}.
}
{By learning survey-specific noise patterns from repeated observations, TSN demonstrates how empirical paired learning can extend the chemical reach of existing spectroscopic data, providing a template applicable to other multi-visit surveys.}

\keywords{Galaxy: evolution -- Galaxy: halo -- Galaxy: disk -- Galaxy: stellar content -- Stars: abundances -- Methods: data analysis}
\maketitle
\nolinenumbers

\section{Introduction}
\label{sec:intro}

High-resolution stellar spectroscopy underpins Galactic archaeology by enabling large-scale inference of stellar parameters and detailed chemical abundances \citep[e.g.][]{2013NewAR..57...80F, 2019ARA&A..57..571J}.
In combination with astrometry and kinematics, such chemo-dynamical data provide constraints on the formation and evolution of the Milky Way disc, bulge, and halo \citep[e.g.][]{2021A&A...656A.156Q, 2024MNRAS.527.1915B}.
To enable population-level studies, modern multiplexed surveys invest heavily in automated pipelines that deliver homogeneous stellar labels \citep[e.g.][]{2017AJ....154...94M,2015MNRAS.449.2604D}.
In APOGEE, the APOGEE Stellar Parameters and Chemical Abundances Pipeline (ASPCAP) infers atmospheric parameters and abundances by fitting synthetic spectral grids \citep{2016AJ....151..144G}.
However, pipeline performance degrades sharply at low signal-to-noise ratios (S/N), where spectral features become difficult to distinguish from instrumental noise and reduction residuals.

Key Galactic components---the distant bulge and inner disc behind high extinction, the outer halo, and satellite systems---are probed primarily by faint targets that yield low S/N spectra.
Recovering chemically informative labels from these observations is therefore critical for studies of bulge and halo formation and for identifying accreted substructure in the Milky Way's outskirts \citep[e.g.][]{2020ApJ...891...39Y, 2023MNRAS.520.5671H}.
Yet APOGEE has accumulated millions of individual visit spectra at low S/N, where single-visit observations are noise dominated and the corresponding co-added ``combined'' spectra, although improved, \modified{do not always reach the high-S/N regime preferred for precise abundance work.}
In this regime, grid-fitting pipelines return imprecise stellar labels with large uncertainties or unreliable flags, and visit spectra are often excluded from abundance-driven analyses.

Nevertheless, recent studies have demonstrated that low-S/N spectra can in fact yield substantial new insights when analysed with advanced machine-learning techniques.
In particular, the application of convolutional neural networks to Gaia RVS spectra has revealed that even at low S/N, accurate stellar parameters can be recovered, significantly enhancing the scientific return of the RVS dataset \citep{2024A&A...682A...9G}.
These advances open the door to the exploitation of large, precise statistical samples, enabling population-level studies that were previously inaccessible, as illustrated by the new results obtained in \citet{2024A&A...688A.167N} and \citet{2025arXiv250706863N}.

\modified{APOGEE is a high-resolution near-infrared spectroscopic survey designed to map the chemo-dynamical structure of the Milky Way using H-band stellar spectra \citep{2013AJ....146...81Z,2017AJ....154...94M}.}
APOGEE provides a unique empirical advantage through its co-added-visit observing strategy.
Many stars have been observed multiple times, and for a subset the combined spectrum reaches high S/N.
These repeated observations define empirical low-/high-S/N pairs for the same star observed with the same instrument, which can be used to supervise denoising without assuming an explicit noise model.
This paired-learning paradigm is closely related to Noise2Noise-style training in imaging \citep{2018arXiv180304189L}, but here the ``twin'' is an actual higher-quality spectrum of the same astrophysical source.
Such empirical spectral twins encode which spectral structures are stable (astrophysical signal) versus those that vary stochastically (instrumental noise and reduction residuals), providing direct supervision for noise suppression that is anchored to survey-specific systematics.

Despite this opportunity, existing approaches to low-S/N label inference have not fully exploited APOGEE's paired structure.
Data-driven label-transfer methods (e.g. \emph{The Cannon}; \citealt{2015ApJ...808...16N}) and deep-learning estimators (e.g. StarNet; \citealt{2018MNRAS.475.2978F}) can infer labels from noisy spectra and have been shown to perform well down to relatively low S/N.
However, purely discriminative spectrum-to-label models do not provide an explicit denoised spectrum for inspection, and can regress toward the training-set mean when the inputs are dominated by measurement noise \citep{2025OJAp....8E..95T}.
Additionally, APOGEE visit spectra can exhibit persistence, detector artefacts, and reduction residuals that vary from visit to visit \citep{2015AJ....150..148H, 2020AJ....160..120J}, introducing correlated structure beyond simple per-pixel Poisson noise that discriminative models may not capture effectively.

A complementary strategy that addresses these limitations is to couple label inference to an empirical spectral reconstruction objective.
Physically motivated forward models such as \emph{The Payne} \citep{2019ApJ...879...69T} predict spectra as a function of labels, enabling residual inspection alongside label inference.
\modified{A data-driven variant, the \emph{dd-Payne} \citep{2019ApJS..245...34X}, replaces the synthetic spectral grid with an emulator trained on observed reference spectra, thereby avoiding biases associated with incomplete or imperfect physical models.}
More recently, hybrid architectures couple label prediction to auxiliary spectral reconstruction \citep[e.g. cross-resolution reconstruction,][]{2024ApJS..272....2L}.
However, these approaches typically rely on synthetic models or cross-dataset mappings that may introduce physical priors not present in the data.
For low-S/N stellar spectra, a practical requirement is therefore to suppress stochastic noise while remaining anchored to an empirically validated reference spectrum from the same survey and a well-defined label scale.
By training on empirical visit--combined spectral twins, such a model can learn survey-specific noise patterns and reduction residuals while preserving APOGEE's internal reference frame.

In this work we introduce TwinSpecNet (TSN), an empirical paired-learning framework for rescuing low-S/N APOGEE visit spectra.
TSN uses a Vision Transformer (ViT) encoder trained sequentially with two objectives: (i) a paired denoising objective that maps low-S/N visits to high-S/N combined spectra, and (ii) a probabilistic label model that predicts stellar parameters and abundances with calibrated uncertainties on the ASPCAP scale.

We assess whether paired learning can extend the effective chemical reach of APOGEE.
Specifically, we test whether TSN (a) improves label precision without introducing large biases relative to ASPCAP, (b) tightens intra-cluster abundance dispersions as an internal consistency check, (c) recovers cleaner chemical sequences in bulge and satellite samples at low S/N, and (d) preserves age-sensitive abundance correlations such as C/N.

Section~\ref{sec:data} defines the paired training samples, label training sample, and preprocessing procedure.
Section~\ref{sec:method} describes the TSN architecture and paired-learning objectives.
Section~\ref{sec:test} validates flux reconstruction, label residuals, and uncertainty calibration on held-out data.
Section~\ref{sec:result} applies TSN to APOGEE science samples.
Finally, Section~\ref{sec:discussion} discusses limitations and broader implications.

\section{Data}
\label{sec:data}
APOGEE employs two cryogenic multi-object spectrographs, each fed by 300 fibres, to obtain spectra with resolving power $R \simeq 22,500$ over the $\unit[1.51 - 1.70]{\mu m}$ wavelength range \citep{2019PASP..131e5001W}.
Stellar parameters and elemental abundances for APOGEE targets are derived with ASPCAP, which estimates stellar labels by minimising the $\chi^2$ distance between observed spectra and a grid of synthetic models \citep{2016AJ....151..144G}.
In its Data Release 17 (DR17) catalogue, ASPCAP provides primary stellar parameters together with multi-element abundances for about $7.3\times10^6$ stars, spanning $T_\mathrm{eff} \in \unit[3500, 7000]{K}$, $\log g \in \unit[0.5, 5.0]{dex}$, and $\feh \in \unit[-2.5, 1.0]{dex}$ with typical precisions of $\simeq 2\%$, \unit[0.1]{dex}, and \unit[0.05]{dex}, respectively.

\modified{We draw two training samples from DR17: a flux-training set of ${\sim}3.2$ million empirical low-/high-S/N visit--combined spectral pairs that supervises the denoising objective (Sect.~\ref{sec:data-pairs}), and a label-training set of $550,354$ giant visit spectra paired with high-quality ASPCAP labels that supervises abundance and parameter estimation (Sect.~\ref{sec:data-parent}). Both samples share a common continuum-normalisation pipeline and are partitioned into training, validation, and test subsets as described in Sects.~\ref{sec:data-preprocess} and \ref{sec:data-splits}.}

\subsection{Flux-training sample: low-S/N visits and high-S/N spectra}
\label{sec:data-pairs}
TSN employs two sequential training objectives---flux reconstruction and label prediction---each requiring distinct supervision signals.
We therefore construct two complementary reference samples from APOGEE DR17.
The first is a purely spectroscopic flux-training set used to learn a survey-specific mapping from low-S/N visits to high-S/N combined spectra.

We build this sample from empirical spectral twins: low-S/N visit spectra paired with high-S/N star spectra of the same star observed with the same telescope, selected from the APOGEE \texttt{allStarLite} catalogue.
For each \texttt{APOGEE\_ID} and telescope we select visit spectra with $10 \leqslant \mathrm{S/N_{visit}} \leqslant 100$ and corresponding combined spectra with $150 \leqslant \mathrm{S/N} \leqslant 300$.
We exclude spectra with STARFLAG bits 0 (bad pixels), 3 (saturation), 9 (persistence), 16 (bright neighbour), or 17 (very bright neighbour) set.
This procedure yields $3,222,879$ candidate visit--star pairs for $141,003$ unique stars.
The number of repeat visits per star spans a wide range (\modified{1 to 2205}, with a mean of 23), providing many empirical low- versus high-S/N pairs and exposing repeatable reduction residuals that a paired model can learn to suppress.
This flux-training sample is purely spectroscopic and does not use ASPCAP labels.

\subsection{Label-training sample: low-S/N visits and ASPCAP labels}
\label{sec:data-parent}
To train the label estimator we apply additional quality cuts to the flux-training sample, yielding a second, more restricted reference set.
In this label-training sample the input remains the low-S/N visit spectrum, but the target is now the ASPCAP label vector of the corresponding highest-S/N combined spectrum rather than the high-S/N flux itself.

\begin{figure}[ht!]
\centering
\includegraphics{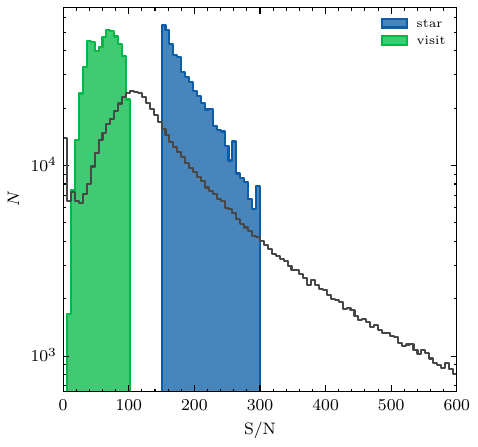}
\caption{
Signal-to-noise distributions for the label-training sample.
Green and blue histograms show the S/N distributions of visit and star (combined) spectra, respectively, from which we adopt the ASPCAP labels for the label-training sample.
The grey curve shows the S/N distribution for the entire ASPCAP DR17, illustrating that low-S/N spectra (S/N$< 100$) comprise roughly 30\% of the total APOGEE population.
}
\label{fig:snr}
\end{figure}

We construct this sample by limiting the surface gravity to $\log g < \unit[3.5]{dex}$ and requiring macroturbulent velocities $v_\mathrm{macro} > \unit[2]{km\,s^{-1}}$ so that the set is dominated by giants.
We exclude stars with STARFLAG bits 0, 3, 9, 16, or 17 set (as above) and ASPCAPFLAG bits 0 (TEFF\_WARN), 3 (LOGG\_WARN), 16 (STAR\_WARN, parameters near grid edge), or 17 (CHI2\_WARN) set.
For each remaining star we retain the signal-to-noise ratio, stellar parameter ($t_\mathrm{eff}$, $\log g$, \feh, $v_\mathrm{macro}$, and microturbulent velocity $v_\mathrm{micro}$) and element-by-element abundances for C, C\,I, N, O, Na, Mg, Al, Si, S, K, Ca, Ti, Cr, and Ni, which we convert to absolute $\xh{X}$ values.
Other elements are discarded because their abundances are too incomplete in this giant-dominated sample.
The final label-training set contains 550,354 low-S/N visit spectra \modified{from $42,572$ unique stars} paired with ASPCAP stellar parameters and abundances from their corresponding high-S/N co-added spectra, with the most metal-poor object having $\feh \sim \unit[-2.3]{dex}$.
\modified{Because this sample is restricted to quality-selected giants, the ASPCAP reference labels are among the most precise in DR17; for $T_\mathrm{eff}$, the mean ASPCAP-reported uncertainty in the label-training sample is around \unit[8]{K}.}
Figure~\ref{fig:snr} shows the S/N distributions of both the low-S/N visit spectra and their paired high-S/N star spectra in this label-training sample.

\subsection{Preprocessing}
\label{sec:data-preprocess}
For both training samples, we work from the APOGEE \texttt{apStar} data products, in which the individual visit spectra and the combined per-star spectra have already been corrected for their derived radial velocities and shifted to the stellar rest frame.

Before entering the network, we continuum normalise each spectrum by fitting a smooth pseudo-continuum independently within each of the three contiguous detector-chip segments \modified{($\unit[{\sim}1.515-1.580]{\mu m}$, $\unit[{\sim}1.586-1.642]{\mu m}$, and $\unit[{\sim}1.648-1.695]{\mu m}$), using a 19-pixel median filter followed by a fifth-order polynomial fit. The continuum estimate is divided out after sigma clipping to suppress emission features while preserving absorption lines.}
We apply the same preprocessing to low-S/N visits and high-S/N reference spectra so that the network learns differences driven by noise and not by normalisation choices.

\subsection{Train/validation/test splits}
\label{sec:data-splits}
For both flux training and label training, we split the visit spectra into training, validation, and test subsets (70:20:10).
Because the label-training data are stored at the per-visit level, this split is performed per visit rather than per \texttt{APOGEE\_ID}, and different  visits of the same star can therefore appear in different subsets.
We thus refer to the evaluation subsets as held-out visit spectra rather than strictly independent stars.

\section{Methods}
\label{sec:method}

\begin{figure}[ht!]
\centering
\includegraphics[width=1\textwidth]{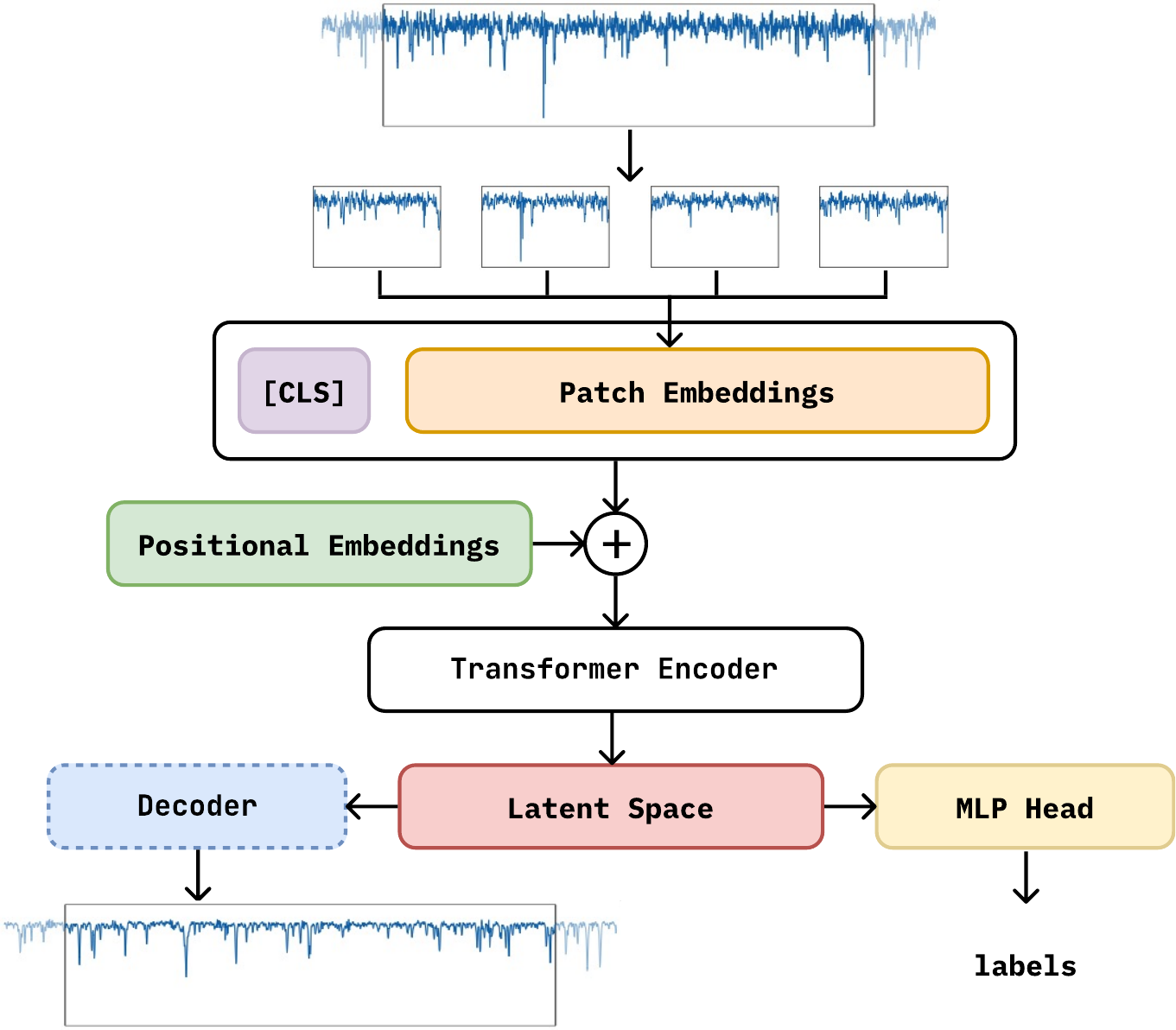}
\caption{
Schematic overview of the TwinSpecNet architecture.
Continuum-normalised APOGEE spectra are \modified{partitioned} into patch tokens and processed by a shared Vision Transformer (ViT) encoder.
The latent representation is then mapped to high-S/N-like flux by a denoising decoder (left) and to stellar labels with uncertainties by an MLP head (right).
In practice we train these objectives sequentially: we pretrain the encoder with the denoising objective and then reuse its weights for label training after replacing the decoder with the probabilistic label head.
}
\label{fig:method}
\end{figure}

\subsection{Transformer architecture}
\label{sec:method-arch}
Fig.~\ref{fig:method} summarises the TwinSpecNet architecture.
TwinSpecNet represents each continuum-normalised APOGEE spectrum as a one-dimensional flux vector and partitions it into 66 fixed-length patches.
We implement the patch embedding as a one-dimensional convolution with kernel size and stride equal to the patch length.
If the input length is not an integer multiple of the patch size, we zero-pad the spectrum at the red end so that all patches have equal length.
A trainable classification token is prepended to the patch sequence, and we add fixed one-dimensional sinusoidal positional encodings to preserve wavelength ordering.

The encoder follows the ViT design, using stacked multi-head self-attention and feed-forward blocks to build a global latent representation that captures both local line structure and long-range correlations across the H band \citep{2017arXiv170603762V,2020arXiv201011929D}.

\subsection{Paired denoising objective}
\label{sec:method-denoise}
We train TSN in two sequential stages: first we pretrain the encoder on the flux reconstruction objective using the larger flux-training set (Section~\ref{sec:data-pairs}), then we fine-tune the encoder for label prediction on the more restrictive labeled set (Section~\ref{sec:data-parent}).
This approach allows the encoder to first learn robust spectral representations from the full diversity of paired observations before specializing to label inference, improving generalization on low-S/N spectra.

 For flux training we attach a lightweight transformer-based decoder to the shared encoder and train them as a denoising autoencoder.
The decoder operates on the same patch sequence, projects tokens back to patch-length flux vectors, and rearranges them into a reconstructed spectrum on the padded wavelength grid.
 We optimise this denoiser using a mean-squared-error loss between the reconstruction and the paired high-S/N star spectrum from the flux-training sample (Section~\ref{sec:data-pairs}), conceptually similar to Noise2Noise-style training on noisy/clean pairs \citep{2018arXiv180304189L}.
This paired objective suppresses noise while retaining astrophysical line structure because the target spectrum is an empirical high-S/N observation of the same star.
Crucially, TSN does not recover information absent in the low-S/N spectrum; rather, it applies learned regularisation that emphasises spectral patterns correlated with high-S/N references and suppresses uncorrelated noise.
In Fig.~\ref{fig:method}, the denoising branch is the left pathway, and the decoder is shown schematically (dashed) because it operates on the same patch-token representation and mirrors the encoder structure.

\subsection{Label inference and uncertainties}
\label{sec:method-label}

For label training we replace the decoder with a lightweight head that predicts, for each label, a mean and a positive variance parameter.
The head is applied to the encoder classification token, yielding a diagonal Gaussian predictive distribution over the label vector.
We train by minimising the Gaussian negative log-likelihood in standardised label space (equivalently, a heteroscedastic squared loss with a $\log \sigma^2$ penalty) on the label-training sample (Section~\ref{sec:data-parent}).
This formulation is standard for heteroscedastic regression and provides per-label predictive uncertainties under a diagonal Gaussian approximation \citep[e.g.][]{2017arXiv170304977K}.
The label models are initialised from the denoising-pretrained encoder weights and are then fine-tuned end-to-end on the label objective.
\modified{To balance network complexity, we split the labels into two models with a shared encoder architecture: one predicts stellar parameters $(T_\mathrm{eff}, \log g, \feh)$ and the other predicts the remaining labels ($v_\mathrm{micro}$, $v_\mathrm{macro}$, and all elemental abundances).
Both models train on the full training set; no spectra are excluded due to missing abundance measurements.
Instead, missing values are handled element-wise during loss computation: only finite label entries contribute to the loss for each training example, so a spectrum with one missing abundance still trains all other labels.
The stellar parameter and velocity labels are complete. For elemental abundances, most labels have fewer than 4\% missing values (e.g.\ Na: 2.2\%, Al: 3.1\%, S: 3.2\%, Cr: 3.3\%), with Ti being the exception at ${\sim}11\%$.}
All reported label uncertainties correspond to the square root of the predicted variances after transforming back to physical units.
\modified{These predictive variances are learned against ASPCAP point-label targets and do not explicitly propagate the catalog ASPCAP label uncertainties during training.
They should therefore be interpreted as internal predictive uncertainties on the ASPCAP reference scale, rather than as a full absolute error budget.
In the current implementation, we assess them empirically by comparing the predicted uncertainties with the residual scatter on held-out data.}

\section{Validation}
\label{sec:test}
TwinSpecNet is designed to map low-S/N spectra onto high-S/N spectra and labels.
In this section we validate that mapping by assessing label residuals and uncertainty calibration on held-out visit spectra.
We further quantify flux reconstruction on held-out visit--star pairs, \modified{including how the denoising pretraining stage depends on the size of the paired flux-training set}, in Appendix~\ref{sec:test-flux}, and check that model sensitivities track known absorption features in Appendix~\ref{sec:test-interpret}.

\begin{figure*}[ht!]
\centering
\includegraphics{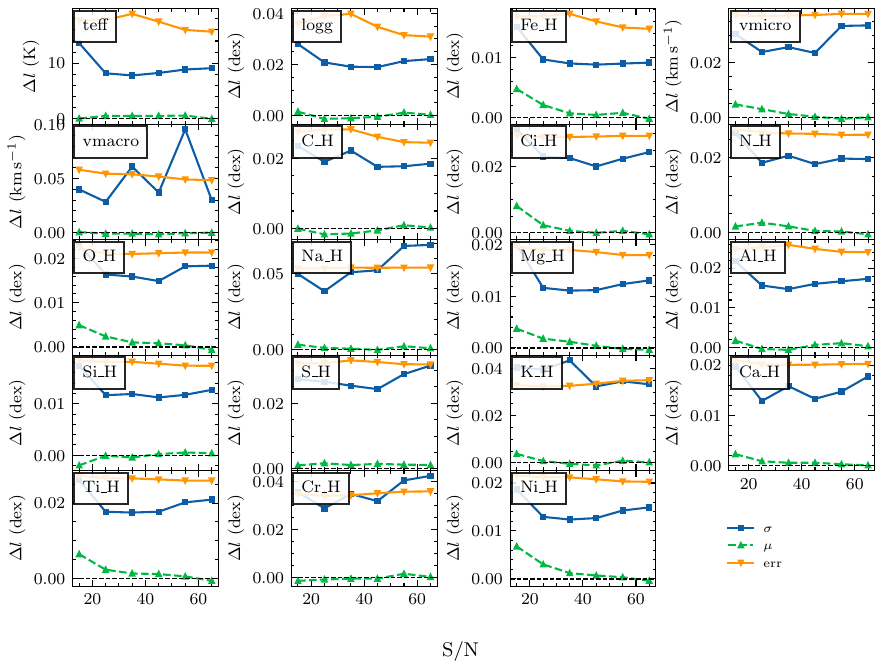}
\caption{
\modified{Label recovery and uncertainty estimates on the test set.}
For each label we show the bias (green triangles), standard deviation (blue squares), and median reported uncertainty \modified{by TSN} (orange circles) as a function of visit $\mathrm{S/N_{visit}}$ on the held-out label-training sample.
\modified{The reported uncertainties broadly track the empirical residual dispersion across S/N, indicating that they provide useful internal predictive uncertainty estimates on the ASPCAP scale.}
\modified{They should not, however, be interpreted as absolute external errors because the systematic uncertainty of the reference labels is not propagated during training.}
}
\label{fig:label_recovery}
\end{figure*}

On the held-out label-training test set, TSN predicts stellar parameters and abundances from low-S/N visits with reduced scatter relative to ASPCAP applied directly to the same visit spectra.
Fig.~\ref{fig:label_recovery} summarises the bias, scatter, and \modified{median TSN-reported uncertainty} as a function of visit $\mathrm{S/N_{visit}}$ from $15$ to $65$ for the held-out label-training sample.
Because the targets are high-S/N ASPCAP labels for quality-selected giants, these comparisons assess internal consistency on the ASPCAP scale rather than absolute external accuracy.

The scatter reduction is largest in the regime where visit-level ASPCAP becomes noise-dominated, with $\mathrm{S/N_{visit}} > 20$ visits showing substantially reduced dispersion for many labels.
Residuals relative to the high-S/N ASPCAP reference show small global biases, with any departures concentrated in the lowest-S/N tail ($\mathrm{S/N_{visit}}\lesssim 15$) where the spectral information content is intrinsically limited.
The scatter and residuals suggest that TSN is capable of reproducing stellar labels even at S/Ns as low as 25, with performance remaining broadly stable across most of the tested low-S/N range.
This consistency aligns with the flux reconstruction results in Fig.~\ref{fig:denoise} and indicates robust performance across the low-S/N tail of the distribution.

We also note that the reported uncertainties are generally larger than or comparable to the internal scatter, which is a direct consequence of the loss function adopted during label inference.
\modified{They are therefore best interpreted as empirically validated internal predictive uncertainties on the ASPCAP scale.}
While these values may still underestimate the full systematic uncertainty (see Section~\ref{sec:result-aspcap}), they provide a useful measure of the internal statistical precision.

\section{Results}
\label{sec:result}
We now evaluate TSN labels in the context of classic APOGEE science applications to assess the empirical gains enabled by paired training.
TSN operates on co-added stellar spectra, but we focus on stars with low combined S/N (below 100) to isolate the regime where standard pipelines often struggle.
The signal-to-noise ratios quoted here refer to the combined spectrum S/N ($\mathrm{S/N}$) for the star.

In this section, we benchmark TSN against ASPCAP (Section~\ref{sec:result-aspcap}) and external high-resolution data (Section~\ref{sec:result-saga}) before examining its influence on open cluster chemistry (Section~\ref{sec:result-cluster}).
We then leverage TSN's improved low-S/N performance to explore the chemical structure of the inner Galaxy (inner-disk and bulge populations; Section~\ref{sec:result-bulge}) and satellite systems (Section~\ref{sec:result-subpop}), probing whether TSN can sharpen constraints on formation scenarios and enrichment histories in these populations.
Finally, we investigate whether TSN preserves and improves age-sensitive C/N diagnostics (Section~\ref{sec:result-age}), which are critical for age-resolved chemo-dynamical studies of the Milky Way.

\modified{All science samples in Sections~\ref{sec:result-aspcap}--\ref{sec:result-age} consist of stars whose combined spectra have $\mathrm{S/N} < 150$. Both training samples---the flux-training set and the label-training set---are constructed exclusively from stars whose combined spectra satisfy $\mathrm{S/N} > 150$ (Sect.~\ref{sec:data}). There is therefore no overlap between the science samples analysed here and the stars used during training, ensuring that the comparisons presented below constitute genuine out-of-sample evaluations.}

\subsection{Comparison against ASPCAP DR17}
\label{sec:result-aspcap}
We first benchmark TSN against ASPCAP DR17 and evaluate performance on \modified{82,402} star spectra with $\mathrm{100 < S/N < 150}$.
Although this is outside the training range ($\mathrm{S/N} \le 100$), it represents a regime where ASPCAP is generally reliable and provides a robust baseline for comparison.
In this high-quality domain, we expect minimal gains from TSN relative to ASPCAP, allowing us to build a realistic estimate of the model's baseline accuracy and precision.

\begin{table}[ht!]
\centering
\caption{Summary of bias ($\mu$), standard deviation ($\sigma$), and MAE for stellar parameters and abundances in comparison with ASPCAP DR17 with $\mathrm{100 < S/N < 150}$. \modified{Here $\mu$ is the mean of $\Delta$, with $\Delta=$ASPCAP$-$TSN.}
\modified{The final column ($\sigma_\mathrm{int}$) gives the APOGEE internal precision from repeat observations of the high-quality giants at APOGEE-2N, taken from Tables~10--11 of \citet{2020AJ....160..120J}.}}
\label{tab:aspcap}
\begin{tabular}{ccrrrr}
\hline
\hline
Parameter & Unit & $\mu$ & $\sigma$ & MAE & $\sigma_\mathrm{int}$ \\
\hline
$T_{\mathrm{eff}}$        & K     & $2.068$ & $18.717$ & $11.567$ & 14 \\
$\log g$                 & dex   & $0.003$ & $0.063$  & $0.041$  & 0.04 \\
$\mathrm{[Fe/H]}$        & dex   & $0.005$ & $0.027$  & $0.015$  & 0.02 \\
$v_{\mathrm{micro}}$     & km\,s$^{-1}$ & $-0.029$ & $0.175$ & $0.095$ &  \\
$v_{\mathrm{macro}}$     & km\,s$^{-1}$ & $0.014$ & $0.157$ & $0.045$ &  \\
$\mathrm{[C/H]}$         & dex   & $<0.001$  & $0.066$  & $0.036$  & 0.03 \\
$\mathrm{[C\,I/H]}$      & dex   & $-0.004$  & $0.074$  & $0.044$  & 0.04 \\
$\mathrm{[N/H]}$         & dex   & $-0.004$  & $0.086$  & $0.038$  & 0.04 \\
$\mathrm{[O/H]}$         & dex   & $-0.001$  & $0.057$  & $0.033$  & 0.04 \\
$\mathrm{[Na/H]}$        & dex   & $-0.017$  & $0.233$  & $0.137$  & 0.12 \\
$\mathrm{[Mg/H]}$        & dex   & $-0.004$  & $0.040$  & $0.022$  & 0.03 \\
$\mathrm{[Al/H]}$        & dex   & $-0.002$  & $0.061$  & $0.040$  & 0.04 \\
$\mathrm{[Si/H]}$        & dex   & $-0.002$  & $0.038$  & $0.023$  & 0.03 \\
$\mathrm{[S/H]}$         & dex   & $0.010$ & $0.135$  & $0.082$  & 0.07 \\
$\mathrm{[K/H]}$         & dex   & $-0.008$  & $0.105$  & $0.064$  & 0.06 \\
$\mathrm{[Ca/H]}$        & dex   & $-0.002$  & $0.060$  & $0.028$  & 0.03 \\
$\mathrm{[Ti/H]}$        & dex   & $0.001$ & $0.077$  & $0.045$  & 0.04 \\
$\mathrm{[Cr/H]}$        & dex   & $-0.001$  & $0.162$  & $0.094$  & 0.07 \\
$\mathrm{[Ni/H]}$        & dex   & $-0.002$  & $0.047$  & $0.028$  & 0.03 \\
\hline
\end{tabular}
\end{table}

Table~\ref{tab:aspcap} presents the bias ($\mu$), standard deviation ($\sigma$), and Mean Absolute Error (MAE) for stellar parameters and abundances in this reference regime, where $\mu$ is computed from $\Delta=\mathrm{ASPCAP}-\mathrm{TSN}$.
\modified{For context, the table also includes the APOGEE internal precision ($\sigma_\mathrm{int}$) derived from repeated observations of the high-quality giants observed at APOGEE-2N \citep[Tables~10--11,][]{2020AJ....160..120J}, which provides a survey-level benchmark for the minimum achievable scatter independent of any modelling assumptions.}
The overall biases are close to zero, though we observe larger scatter ($\sigma>0.1$) for Na, Cr, S, and K, which are known to be difficult elements in APOGEE due to weak lines or telluric contamination \citep{2020AJ....160..120J}.
For other labels, the MAE is comparable to the typical uncertainties reported by TSN (usually within a factor of 1--2), suggesting that our error estimates are reasonable proxies for the internal precision.
For $T_\mathrm{eff}$, the measured TSN scatter of \unit[18.7]{K} is larger than the APOGEE repeat-based giant precision of \unit[14]{K}, so it does not imply a precision better than the empirical ASPCAP floor.
\modified{For reference, the mean formal ASPCAP uncertainty in $T_\mathrm{eff}$ for the label-training sample is \unit[8.21]{K}, but we regard the repeat-based precision as the more empirical benchmark.}

In this intermediate-S/N regime, TSN predictions track ASPCAP values with small biases and scatters comparable to the internal precision of the survey, confirming that TSN successfully reproduces the reference label scale when the input data quality is moderate.
These metrics also provide a reliable estimate of the internal accuracy across the low-S/N tail, because the performance of TSN is largely independent of SNR (Section~\ref{sec:test}).

\begin{figure}[ht]
\centering
\includegraphics{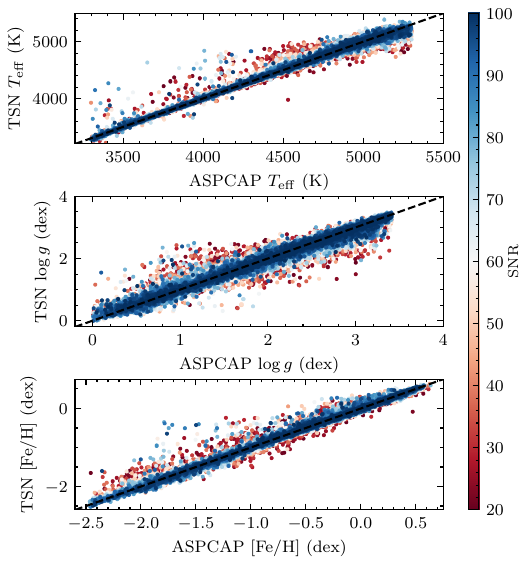}
\caption{
Comparison of TSN and ASPCAP parameters at low S/N.
The top, middle, and bottom panels show $T_\mathrm{eff}$, $\log g$, and $\mathrm{[Fe/H]}$, respectively.
Each panel compares the TSN label to the ASPCAP reference label for stars with $\mathrm{S/N_{visit}}<100$, with points colour-coded by visit $\mathrm{S/N_{visit}}$.
ASPCAP should be interpreted here as a reference label scale rather than as ground truth.
}
\label{fig:residual}
\end{figure}

We then focus on the low-S/N regime by comparing predictions for stars with $\mathrm{S/N_{visit}}<100$.
Fig.~\ref{fig:residual} compares TSN and ASPCAP labels for $T_\mathrm{eff}$, $\log g$, and $\mathrm{[Fe/H]}$, colour-coded by S/N.
TSN follows the ASPCAP reference scale closely, with small global biases ($\unit[2.3]{K}$ in $T_\mathrm{eff}$ and \modified{$<\unit[0.01]{dex}$ in $\log g$ and $\feh$}) and scatters of $\unit[35.0]{K}$, $\unit[0.10]{dex}$, and $\unit[0.05]{dex}$, respectively.
Compared to the values in Table~\ref{tab:aspcap}, the larger deviations are concentrated in the lowest-S/N tail ($\mathrm{S/N_{visit}}\lesssim 60$), where the information content of the spectrum is most degraded.
Since ASPCAP itself is not ground truth at low S/N, this comparison primarily verifies that TSN remains anchored to the reference scale and does not introduce large spurious offsets in the noise-dominated regime.

\subsection{Comparison against SAGA}
\label{sec:result-saga}

\begin{figure}[ht!]
\centering
\includegraphics{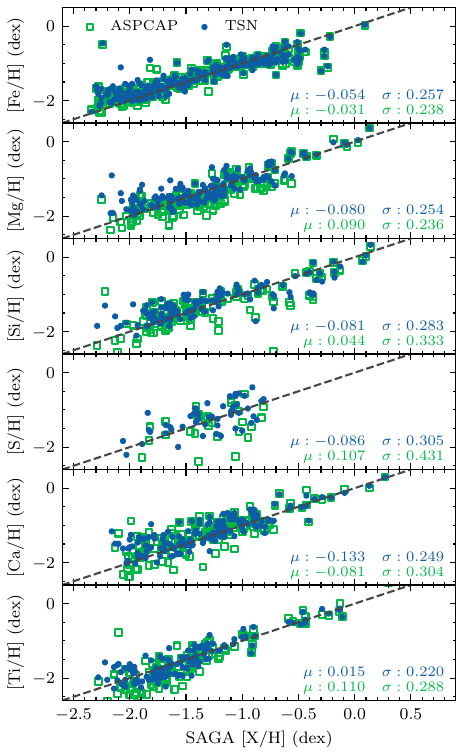}
\caption{
Comparison of TSN and ASPCAP abundances against high-resolution literature values from the SAGA database.
The horizontal axis shows the SAGA abundance, while the vertical axis displays the difference (Model $-$ SAGA) for Fe, Mg, Si, S, Ca, and Ti (top to bottom).
Blue circles and green open circles represent predictions from TSN and ASPCAP, respectively.
The bias and standard deviation \modified{in \unit{dex}} for each method are shown in the top-right corner of each panel, \modified{with $\mu$ computed from $\Delta=\mathrm{SAGA}-\mathrm{Model}$}.
The sample is restricted to stars with at least 50 matches and has a median $\mathrm{S/N} \sim 60$.
}
\label{fig:saga}
\end{figure}

To assess accuracy against an external scale, we cross-match TSN predictions with the Stellar Abundances for Galactic Archaeology (SAGA) database \citep{2008PASJ...60.1159S}, a compilation of high-resolution abundances for metal-poor stars.
This comparison provides a stress test of the model at the metal-poor edge of our training set, where label recovery is most challenging.

We restrict the comparison to stars with $\feh > \unit[-2.3]{dex}$ (the limit of our training set) and to elements with at least 50 common stars.
\modified{This yields 265 stars for \feh, with smaller subsets for individual abundances (\xh{Mg}: 159, \xh{Si}: 125, \xh{Ca}: 167, \xh{Ti}: 131, \xh{S}: 50).}
This sample has median $\mathrm{S/N}\sim 60$, a good place to test the improvement from TSN.
TSN recovers the one-to-one relation with literature abundances, yielding results for [Fe/H] \modified{and [Mg/H]} that are nearly identical to that from ASPCAP.
For $\alpha$-elements such as S, Ca, and Ti, TSN reduces the scatter and improves bias estimates relative to ASPCAP, as shown in Fig.~\ref{fig:saga}.
This confirms that TSN's noise suppression does not distort the fundamental abundance scale relative to independent high-resolution measurements, at least within the metallicity range covered by the training data.

\subsection{Cluster chemical homogeneity}
\label{sec:result-cluster}
\begin{figure}[ht!]
\centering
\includegraphics{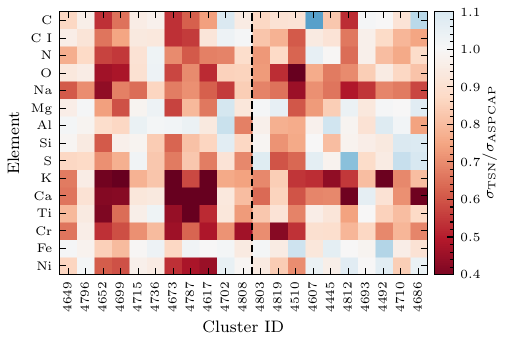}
\caption{
Intra-cluster abundance dispersion ratios for APOGEE cluster members from \citep{2024A&A...686A..42H}.
For each cluster and element, the ratio of the abundance dispersion derived from TSN to that from ASPCAP, $\sigma_\mathrm{TSN}/\sigma_\mathrm{ASPCAP}$, is computed.
This ratio quantifies the relative precision of TSN; values below unity indicate that TSN yields tighter sequences than ASPCAP.
Clusters are ordered by the number of stars passing the selection, and only clusters with at least 25 members are shown.
The left and right blocks correspond to a low-S/N-only selection ($\mathrm{S/N_{visit}}<100$) and a mixed selection ($\mathrm{S/N_{visit}}<150$) used to reach the membership threshold.
Ratios below unity indicate reduced noise-driven scatter in TSN labels in the low-S/N regime, while ratios near unity for well-measured elements (e.g. Fe, Mg, Si) indicate that TSN does not artificially tighten already precise labels.
}
\label{fig:cluster}
\end{figure}

Star clusters provide a stringent internal consistency check because their member stars are expected to have small intrinsic abundance dispersions for many elements \citep[e.g.][]{2006AJ....131..455D}.
We cross-match APOGEE DR17 stars to the cluster catalogue of \citep{2024A&A...686A..42H} and compute intra-cluster dispersions for TSN and ASPCAP labels, \modified{yielding a total of 2,569 cluster member stars}.
To isolate the low-S/N regime we consider a sample with $\mathrm{S/N}<100$.
It is expanded to $\mathrm{S/N}<150$ when needed to reach a minimum membership of 25 stars.

Fig.~\ref{fig:cluster} displays the dispersion ratio $\sigma_\mathrm{TSN}/\sigma_\mathrm{ASPCAP}$ for a range of elements, with clusters ordered by the number of member stars.
The left panel shows results for the strict low-S/N selection ($\mathrm{S/N}<100$), while the right panel includes stars up to $\mathrm{S/N}<150$ to increase the sample size.
The ratio is typically below unity across most elements for low-S/N-dominated clusters, indicating reduced noise-driven scatter in TSN labels.
For elements that are already precise in ASPCAP (e.g. Fe, Mg, Si), ratios remain near unity, consistent with TSN improving noisy measurements without over-tightening well-measured labels.

As a failure mode, a model that globally compresses abundance space would also reduce dispersions in the mixed selection and erase cluster-to-cluster offsets, which we do not observe.
To verify that TSN does not reduce scatter by globally compressing abundance space, we decomposed the total variance into within-cluster and between-cluster components.
TSN reduces within-cluster variance at low S/N (median reduction of 40\% for all elements) while preserving between-group variance to within 5\% (i.e., cluster-to-cluster offsets change by $<$5\% on average), demonstrating that the improved precision is not driven by regression toward a global manifold.

\subsection{Bulge and inner-disk chemical structure}
\label{sec:result-bulge}
The inner Galaxy contains overlapping bulge and inner-disk populations with distinct $\alpha$-element sequences and metallicity distributions \citep{2008A&A...486..177Z,2017A&A...605A..89B, 2021A&A...656A.156Q, 2025arXiv250706863N}, offering critical constraints on Milky Way formation scenarios.
These populations are often sampled at low S/N, where abundance uncertainties have historically limited population decomposition and chemical-evolution modelling.
Here we investigate whether TSN's noise suppression sharpens chemical structure in two complementary selections: a geometric inner-Galaxy sample that is dominated by the inner disk, and a dedicated bulge subset from the RPM (reduced proper motion) sample of \citet{2025arXiv250706863N}.
The RPM approach uses near-infrared photometry plus proper motions to build a pseudo-absolute-magnitude diagnostic that separates foreground disc contaminants from stars concentrated toward the bulge-bar region \citep{2021A&A...656A.156Q}.
In \citet{2021A&A...656A.156Q}, the cleaned sample is constructed by applying a box-like cut in the $(J-K_s)_0$--RPM diagram that targets the innermost overdensity while suppressing foreground red-clump contamination without imposing chemistry-based pre-selection.
This procedure reduces the parent geometric inner-Galaxy sample (about $2.65\times10^4$ stars) to an RPM subset of about $8\times10^3$ stars that is more homogeneous for bulge-bar chemo-dynamical analysis \citep{2021A&A...656A.156Q}.

We first define a geometric inner-Galaxy selection using Galactocentric coordinates computed from Gaia EDR3 photogeometric distances and APOGEE $(l,b)$, requiring a Galactocentric cylindrical radius $R_\mathrm{GC}<\unit[3]{kpc}$ and $|Z|<\unit[5]{kpc}$, following the approach of \citet{2021ApJ...909...77G}.
We apply standard APOGEE quality cuts \modified{by excluding spectra with STARFLAG bits 0 (bad pixels), 3 (saturation), 9 (persistence), 16 (bright neighbour), or 17 (very bright neighbour) set, and ASPCAPFLAG bits 0 ($T_\mathrm{eff}$ warning), 3 ($\log g$ warning), 16 (parameters near grid edge), or 17 (high $\chi^2$) set}, and we require $\xfe{Mg}$ and $\feh$ abundance flags equal to zero.
We then restrict to the spectra with $\mathrm{S/N}<100$.
This yields 8,143 low-S/N inner-Galaxy stars, of which 4,957 are classified as high-$\alpha$ population following the criteria in \citet{2021ApJ...909...77G}.
In parallel, we analyse the RPM-based bulge selection from \citet{2025arXiv250706863N}, which starts from the RPM catalogue (8,061 stars) of \citet{2021A&A...656A.156Q} and keeps stars confined to the inner Galaxy with apocentre $<\unit[5]{kpc}$ (6,211 stars).
\citet{2025arXiv250706863N} then apply a chemo-orbital classification by identifying bar-supporting stars from orbital-frequency criteria, defining non-bar stars with $P_\mathrm{bar}=0$, and splitting this non-bar component by net rotation $L_{z,i}$.
In that scheme, stars with $|L_{z,i}|<0.4$ define the non-rotating spheroidal-bulge component, while stars with $L_{z,i}>0.8$ trace the inner-thick-disc component.
Relative to a purely geometric cut, this sequence reduces thin-disc and halo pass-through contamination and yields a cleaner bona fide bulge sample for chemical interpretation.
After applying our APOGEE quality filters and TSN availability constraints to this RPM-based bulge selection, the final sample used in this section contains 692 stars, of which 274 have $\mathrm{S/N}<100$ (see Chiappini et al. in prep for a full analysis of the bona fide bulge stars).

\begin{figure*}[ht!]
\centering
\includegraphics[width=0.88\textwidth]{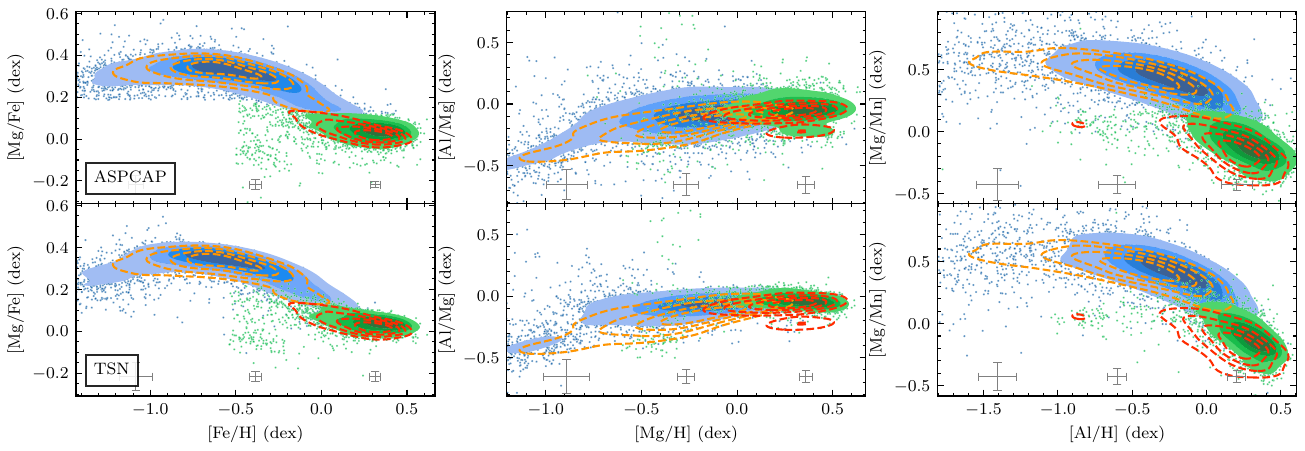}
\caption{
Inner-disk abundance-plane comparison at low S/N.
Panels show $\xfe{Mg}$--$\feh$ (left), $\mathrm{[Al/Mg]}$--$\xh{Mg}$ (middle), and $\mathrm{[Mg/Mn]}$--$\xh{Al}$ for stars in the inner-Galaxy with $\mathrm{S/N}<100$.
The top and bottom rows compare ASPCAP and TSN labels for the same stars.
Points are coloured by the $\alpha$-rich/$\alpha$-poor split defined in the ASPCAP $\xfe{Mg}$--$\feh$ plane.
Contours trace the low-S/N density (levels 0.2, 0.4, 0.6, 0.8) and are compared to a high-S/N reference sample (combined spectra with $\mathrm{S/N}>150$, shown as dashed contours).
The typical uncertainty in the abundances is shown as a function of metallicity or [Mg/H] across the bottom of each panel.
TSN reduces noise-driven broadening and suppresses the low-S/N Al/Mg artefact (the ``Al finger''), bringing the low-S/N distribution closer to the high-S/N reference.
We adopt the Mn abundance from ASPCAP.
}
\label{fig:inner}
\end{figure*}

\begin{figure*}[ht!]
\centering
\includegraphics{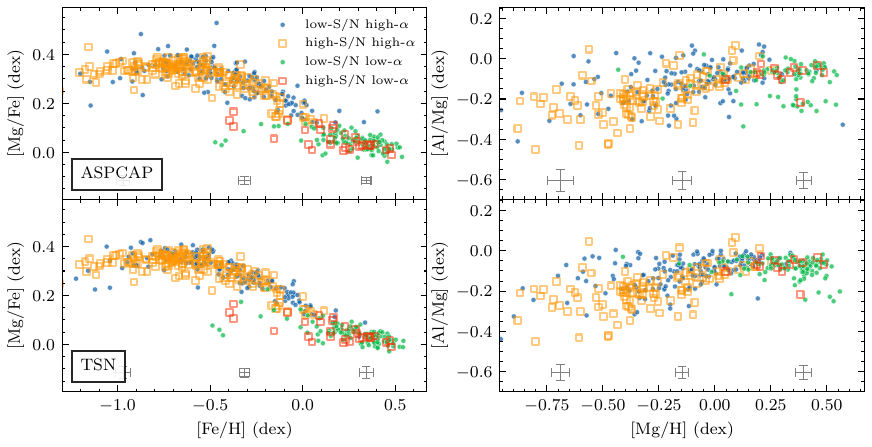}
\caption{
Bulge abundance-plane comparison at low S/N.
Panels show $\xfe{Mg}$--$\feh$ (left) and $\mathrm{[Al/Mg]}$--$\xh{Mg}$ (right) for stars in the RPM-bulge selections with $\mathrm{S/N}<100$.
The top and bottom rows compare ASPCAP and TSN labels for the same stars.
Points are coloured by the $\alpha$-rich/$\alpha$-poor split defined in the ASPCAP $\xfe{Mg}$--$\feh$ plane, following the same colour scheme as in Fig.~\ref{fig:inner}.
The hollow squares represent the high-S/N reference sample (combined spectra with $\mathrm{S/N}>150$).
The typical uncertainty in the abundances is shown as a function of metallicity or [Mg/H] across the bottom of each panel.
In the $\mathrm{[Al/Mg]}$--$\xh{Mg}$ panel, TSN includes 31 more stars than ASPCAP because those stars are flagged for Al abundance in ASPCAP.
}
\label{fig:bulge}
\end{figure*}

Fig.~\ref{fig:inner} compares the low-S/N inner-Galaxy sample in the $\xfe{Mg}$--$\feh$ and $\mathrm{[Al/Mg]}$--$\xh{Mg}$ planes, split into $\alpha$-rich and $\alpha$-poor sequences using the ASPCAP $\xfe{Mg}$--$\feh$ plane.
In Fig.~\ref{fig:inner}, we also include the $\mathrm{[Mg/Mn]}$--$\mathrm{[Al/Fe]}$ plane, which is widely used for chemical diagnostics \citep[e.g.][]{2020MNRAS.493.5195D, 2023MNRAS.519.3611F, 2026arXiv260212415E}.
The low-S/N sample is shown as filled contours for ASPCAP (top row) and TSN (bottom row), with the high-S/N reference population shown as dashed contours.
Typical abundance uncertainties are shown at the bottom of each panel as a function of the horizontal axis.

In $\xfe{Mg}$--$\feh$, the TSN low-S/N contours closely match the high-S/N reference, indicating that TSN suppresses noise-driven broadening without distorting the global chemical structure.
By reducing this systematic effect, TSN enables cleaner separation of true chemical enrichment patterns from measurement noise, strengthening constraints on inner-disk and bulge formation scenarios and the relative importance of different stellar populations.

In the Al-based plane, TSN substantially reduces the spurious low-S/N ``Al finger'' at \xfe{Al} $\sim \unit[0.2]{dex}$ \citep[see][]{2020AJ....160..120J} and tightens the distribution towards the high-S/N locus.
The suppression of the Al finger is particularly significant because this feature is a known artefact in APOGEE that can mimic Al enhancement and complicate the interpretation of inner-Galaxy chemistry \citep{2020AJ....160..120J}.
However, TSN's Al predictions are conditioned on the label-training reference set, which adopts ASPCAP labels only for stars with high-S/N combined spectra ($150 \le \mathrm{S/N} \le 300$; Section~\ref{sec:data-parent}).
Stars with abnormal Al abundances are therefore under-represented in training as they preferentially lie at $\mathrm{S/N}<100$ in APOGEE DR17, and TSN is not expected to recover such rare populations outside its label-training domain.

As for Mn, because it is not included in TSN, we adopt ASPCAP abundances and therefore do not see a difference in derived $\mathrm{[Mg/Mn]}$.

In Fig.~\ref{fig:bulge}, we show the same abundance-plane comparison for the RPM-based bulge sample, displayed as a scatter plot because of the smaller sample size.
The S/N and $\alpha$-sequence selections are matched to those used for the inner-disk analysis.
In this bulge-focused sample, TSN follows the high-S/N reference trends (open squares) more closely than ASPCAP.
In the $\xfe{Mg}$--$\feh$ plane, TSN shows fewer high-$\xfe{Mg}$ outliers (above $\unit[0.45]{dex}$), indicating reduced noise-driven broadening.
The difference is even clearer in the $\mathrm{[Al/Mg]}$ plane: ASPCAP shows a broader low-$\alpha$ distribution and a visible low-S/N Al-finger-like feature, while TSN yields a tighter sequence for both low- and high-$\alpha$ stars.
TSN also reduces the number of stars with anomalously low Al abundance around $\mathrm{[Al/Mg]}\sim\unit[-0.2]{dex}$ at $\xh{Mg}>\unit[0.0]{dex}$.
Such bimodality is not limited to Al, but also found in Cr, Co and Ni \citep{2020AJ....160..120J}.
It is likely an artifact due to improper analysis as this feature is not detected in the high-quality sample as in \citet{2026arXiv260212415E}.
Moreover, TSN provides Al abundances for 31 spectra that are flagged in ASPCAP.
This recovery is important for bulge science because chemically peculiar stars are rare but highly informative for constraining early enrichment channels and population mixing in the inner Galaxy, including stars near the metal-poor tail \citep{2025A&A...704A..44M} and near the boundaries of the label-training domain.

\subsection{Substructure chemical tagging in satellite systems}
\label{sec:result-subpop}
\begin{figure*}[ht!]
\centering
\includegraphics[width=0.88\textwidth]{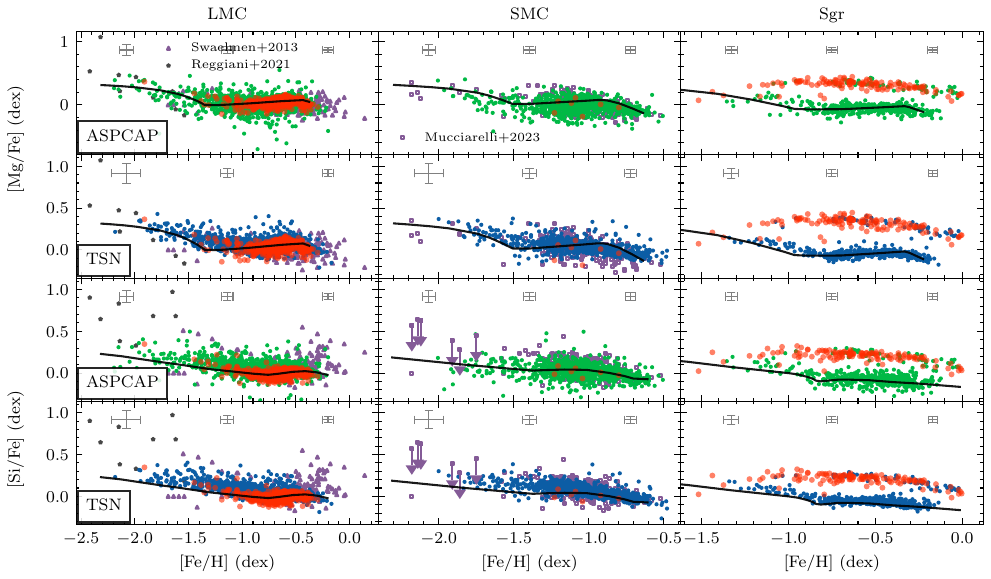}
\caption{
Satellite-system abundance patterns.
Shown are $\xfe{Mg}$--$\feh$ and $\xfe{Si}$--$\feh$ for stars associated with the LMC, SMC, and Sagittarius, comparing ASPCAP (odd rows) and TSN (even rows).
High-S/N reference stars ($\mathrm{S/N}>150$, shown in red) are overplotted, together with reference chemical-evolution tracks from \citet{2021ApJ...923..172H} (solid lines) and literature measurements for metal-poor Magellanic Cloud stars \citep{2013A&A...560A..44V, 2021AJ....162..229R,2023A&A...671A.124M}.
The typical uncertainty in the abundances is shown as a function of metallicity across the top of each panel.
The figure shows ASPCAP predictions in green (odd rows) and TSN predictions in blue (even rows), with low-S/N samples shown as density distributions.
TSN yields tighter low-S/N sequences and reduces noise-driven outliers while preserving the loci traced by high-S/N samples.
The chemical-evolution tracks are included as qualitative guidance and are not an independent validation because they are calibrated on ASPCAP abundances.
}
\label{fig:subpop}
\end{figure*}

Satellite galaxies such as the Large Magellanic Cloud (LMC), Small Magellanic Cloud (SMC) \citep{2020ApJ...895...88N}, and Sagittarius \citep{2017ApJ...845..162H} preserve distinct chemical-enrichment histories that encode their star formation efficiencies, gas retention, and interaction timescales with the Milky Way.
Extending chemical tagging to their fainter members requires robust abundance measurements at low S/N, where distance and target faintness push many stars beyond the reach of standard pipelines.
We investigate whether TSN can recover cleaner abundance sequences in these satellite systems, sharpening constraints on their chemical-evolution pathways and improving membership discrimination.

We select candidate members using simple sky, velocity, Gaia proper-motion, and near-infrared colour--magnitude cuts designed to isolate each system while keeping the selection reproducible, following the selection criteria as in \citet{2023MNRAS.519.3611F}.
For these tests we prepare low-S/N ($\mathrm{S/N}<100$) and high-S/N reference ($\mathrm{S/N}>150$) samples using standard APOGEE quality cuts similar to those applied in the bulge analysis (Section~\ref{sec:result-bulge}).
The resulting low-S/N samples contain 1,652 (LMC), 1,024 (SMC), and 492 (Sagittarius) stars.

Fig.~\ref{fig:subpop} presents $\xfe{Mg}$ and $\xfe{Si}$ versus $\feh$ for each satellite, with ASPCAP (green, odd rows) and TSN (blue, even rows) predictions shown alongside high-S/N reference stars (red) and flexCE chemical-evolution model tracks from \citet{2021ApJ...923..172H}.
The flexCE code \citep{2017ApJ...835..224A} models the chemical evolution of these systems by parametrizing star formation history, gas inflow and outflow rates, and initial mass function-weighted yields, constrained by observed stellar abundances and star formation histories.
In each system, TSN recovers the characteristic low-$\alpha$ abundance patterns seen in high-S/N stars, but with cleaner sequences and fewer noise-driven outliers in the low-S/N regime.
TSN shows reduced scatter in $\xfe{Mg}$ and $\xfe{Si}$ at fixed $\feh$ relative to ASPCAP, improving consistency with the high-S/N loci and with independent literature measurements where available.
The model tracks successfully bracket the observed sequences for both TSN and ASPCAP, but TSN's tighter low-S/N distributions enable sharper constraints on chemical-evolution parameters.

Excessive regularisation that forces stars onto dominant Milky Way abundance trends would pose a significant risk for TSN when applied to satellite systems with distinct chemical enrichment histories.
However, the system-specific offsets and turnover features characteristic of each satellite are preserved in TSN predictions, indicating that the paired-learning approach does not erase population-level abundance structure.
These results indicate that TSN can extend chemically informative samples to fainter members and lower surface densities in external systems while retaining the detailed abundance patterns needed for chemical-evolution modelling and kinematic decomposition.

\subsection{Age information from C/N}
\label{sec:result-age}
\begin{figure}[ht!]
\centering
\includegraphics{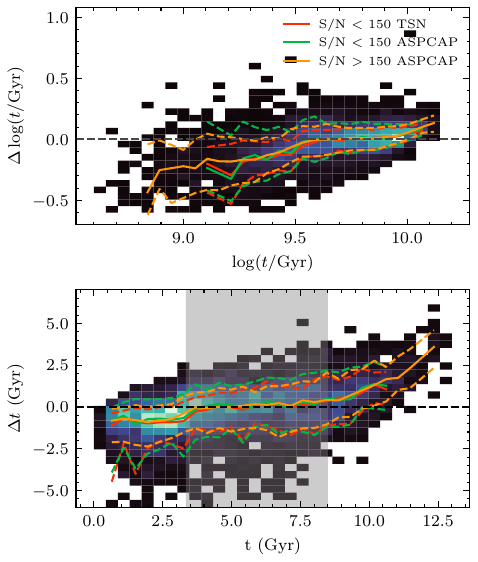}
\caption{
C/N-based ages versus asteroseismic ages.
The horizontal axis shows APOKASC-3 asteroseismic ages \citep{2025ApJS..276...69P}, while the vertical axis shows C/N-based ages inferred from TSN (red) and ASPCAP (green) abundances for low-S/N stars.
Ages are estimated from the empirical [C/N]--age calibration of \citet{2025arXiv250925321R}, which provides a metallicity-dependent mapping from post-dredge-up C/N ratios to stellar ages for red giants.
The high-S/N ASPCAP sample is shown as a background density, and the relation is plotted in both logarithmic and linear scaling.
The APOKASC-3 quality-selected sample contains 8,104 stars, of which 2,767 have $\mathrm{S/N}<150$.
Within the $3.34$--$\unit[8]{Gyr}$ range, where the calibration shows no strong residual bias, the age scatter is $\sim\unit[1.49]{Gyr}$ for low-S/N TSN versus $\sim\unit[1.70]{Gyr}$ for low-S/N ASPCAP.
This demonstrates that paired training preserves age-sensitive abundance information and improves age precision in the low-S/N regime.
}
\label{fig:cn_age}
\end{figure}
We next investigate whether TSN preserves age-sensitive information by combining C and N abundances with APOKASC-3 asteroseismic ages \citep{2025ApJS..276...69P}.
The carbon-to-nitrogen (C/N) ratio in evolved red giants provides a powerful empirical age indicator because it reflects the initial stellar mass (and hence age) through first dredge-up mixing, with additional sensitivity to evolutionary state and metallicity \citep{2002PASP..114..375S,2016MNRAS.456.3655M,2021A&A...654A..13L,2021A&A...645A..85M,2025A&A...703A...4T}.
This age proxy has enabled large-scale age-resolved studies of Galactic structure and chemical evolution \citep{2015MNRAS.453.1855M,2016ApJ...823..114N,2021A&A...654A..13L,2021A&A...645A..85M, 2024A&A...683A.111G, 2025MNRAS.541.2631C}, making it critical to verify that TSN's noise suppression preserves rather than distorts these age-sensitive abundance patterns.

For red giants, the post-dredge-up C/N ratio correlates with mass and hence age, with a metallicity dependence \citep{2016MNRAS.456.3655M}.
Following the empirical [C/N]--age calibration of \citet{2025arXiv250925321R}, we map $\mathrm{[C/N]}$ and $\feh$ to $\log t$ \modified{using their Eq.~4, where $\mathrm{[C/N]} = \xh{C} - \xh{N}$ is computed from either ASPCAP or TSN-predicted $\xh{C}$ and $\xh{N}$, combined with the corresponding $\feh$ estimate.}
Their method treats RGB and red-clump stars separately by applying a metallicity-dependent correction to the red-clump ages; we adopt their APOKASC-3 quality cuts (including seismic $\log g$ consistency, a minimum C/N, and finite age uncertainties).
C/N-based age proxies from TSN and ASPCAP are then compared to the seismic ages for stars with low-S/N spectra, restricting to $\mathrm{S/N}<150$.

Fig.~\ref{fig:cn_age} compares C/N-derived ages to APOKASC-3 asteroseismic ages for the low-S/N sample, with TSN and ASPCAP abundances shown in red and green, respectively, relative to the high-S/N reference distribution. \modified{Unlike the label-recovery figures, we retain this age--age presentation to facilitate direct comparison with the empirical C/N--age calibration of \citet{2025arXiv250925321R}.}
Within the $3.34$--$\unit[8]{Gyr}$ regime where this proxy is most stable, TSN reduces the age scatter from $\sim\unit[1.70]{Gyr}$ (low-S/N ASPCAP) to $\sim\unit[1.49]{Gyr}$, approaching the high-S/N reference dispersion of $\sim\unit[1.61]{Gyr}$.
Despite a slightly larger bias toward younger stars for TSN, the overall mean absolute error decreases from $\unit[1.31]{Gyr}$ (ASPCAP) to $\unit[1.09]{Gyr}$ (TSN), demonstrating improved age precision across the full age range.
\modified{That TSN's scatter marginally undershoots the high-S/N ASPCAP reference is notable, and a plausible explanation lies in known ASPCAP systematics.
\citet{2020AJ....160..120J} document a spurious enhancement of C abundances in cool giants ($T_\mathrm{eff}<\unit[4000]{K}$, the ``C finger'') and an anomalous rise of C/N near the tip of the RGB ($\log g< \unit[1]{dex}$); these biases are most severe at low S/N but are not absent at high S/N.
Through the C/N--age relation, even modest abundance offsets propagate into age residuals, inflating the reference scatter for affected stars.
TSN, trained to reproduce the spectral response of high-S/N data (Appendix~\ref{sec:test-interpret}), anchors its predictions to the cleaner high-S/N signal and may therefore yield a tighter C/N-based age relation for low-S/N stars than ASPCAP achieves at the same S/N.
An analogous suppression of a reproducible ASPCAP artefact is seen in the [Al/Mg] plane (Fig.~\ref{fig:inner}), where TSN attenuates the low-S/N ``Al finger'' that likewise reflects a systematic rather than purely noise-driven effect.}

This improvement is modest relative to the calibration uncertainties in the empirical C/N--age relation itself \citep{2025arXiv250925321R}, which is derived from APOGEE abundances and thus shares systematics with TSN.
At fixed seismic age, TSN reduces noise-driven scatter in C and N without introducing strong low-S/N biases, indicating that the paired objective preserves correlated abundance patterns needed for C/N-based age diagnostics.

Recent work by \citet{2025MNRAS.541.2631C}, based on new high-resolution observations of Kepler giants, demonstrated tighter age--abundance trends than \citet{2025arXiv250925321R}, with $\mathrm{[Ce/Mg]}$ and $\mathrm{[Zr/Ti]}$ emerging as their tightest clocks.
For our validation, the main limitation is that their APOGEE comparison sample consists mostly of nearby stars with high-quality APOGEE spectra (typically $\mathrm{S/N}>200$), leaving no overlap with our low-S/N sample and preventing a direct TSN test on their golden set.
An alternative validation route is to compare the tightness of abundance--age relations against asteroseismic ages, either in APOKASC-3 as in \citet{2025arXiv250925321R} or in the larger 17{,}000-star catalogue of \citet{2026arXiv260206870W}, where ages are inferred with a comparable seismic framework.
In practice, this test remains limited because the relation from \citet{2025MNRAS.541.2631C} that is most applicable to APOGEE data is $\mathrm{[Ce/Mg]}$, but Ce is not included in TSN due to incomplete label coverage, so the observed $\mathrm{[Ce/Mg]}$ scatter is dominated by ASPCAP Ce uncertainties, as seen in Fig.~\ref{fig:inner} for Mn abundance.
We therefore do not obtain a meaningful TSN-versus-ASPCAP conclusion from this specific clock at present, although a broader comparison in \citet{2026arXiv260206870W} using other abundance combinations is planned.

Because APOGEE's low-S/N spectra disproportionately target faint and distant populations, extending robust age information into this regime is valuable for time-resolved Galactic archaeology.
Precise ages link present-day chemo-kinematic structure to the time sequence of disc assembly, and low-mass red giants are especially powerful tracers because they are intrinsically luminous and long-lived, sampling several kpc and more than $\sim 10$~Gyr of Milky Way history.
Asteroseismic constraints provide some of the most direct and homogeneous age estimates for red giants and have been widely used in the Galactic disc \citep[e.g.][]{2013EPJWC..4303004M,2016MNRAS.455..987C,2017A&A...600A..70A,2018MNRAS.475.5487S,2019MNRAS.490.4465R,2019MNRAS.490.5335S,2021A&A...645A..85M,2023MNRAS.524.1634S,2023MNRAS.526.2141W,2024AJ....167...50S,2024A&A...685A.150V,2024AJ....167..208W} and the halo \citep[e.g.][]{2019A&A...627A.173V,2020NatAs...4..382C,2021ApJ...916...88G,2021ApJ...912...72M,2021NatAs...5..640M,2022MNRAS.514.2527B}.
Such ages enable time-resolved interpretation of low-$\alpha$/high-$\alpha$ trends \citep{2021A&A...645A..85M,2021A&A...654A..13L,2023ApJ...954..124I}, age--metallicity relations and abundance gradients, and constraints on radial migration \citep{2025PASJ...77..916B} and on reconstructions of the Galactic star-formation history \citep{2022Natur.603..599X,2024A&A...688A.167N,2025ApJS..280...13W,2025ApJ...994..126B}.
Asteroseismic ages are also used to validate other age-estimation techniques (e.g. chemical clocks and gyrochronology) \citep[e.g.][]{2021NatAs...5..707H,2021A&A...646A..78M,2022A&A...660A..15M} and as training data for machine-learning age inference \citep{2019MNRAS.489..176M,2021MNRAS.503.2814C,2023A&A...678A.158A,2023MNRAS.522.4577L}, but these approaches depend on high-quality training labels and generally become unreliable when extrapolating beyond the training set.
In this paper, our goal is to increase the number of faint APOGEE red giants with chemically driven age information by improving abundance precision at low S/N; even a moderate gain in low-S/N age precision can therefore be scientifically valuable \citep{2024A&A...683A.111G}.

\section{Discussion}
\label{sec:discussion}
The results in Sections~\ref{sec:test} and \ref{sec:result} show that TwinSpecNet does more than denoise spectra: it expands the regime in which APOGEE low-S/N spectra can be treated as chemically informative data.
Here we interpret why paired training produces these gains, and place TSN in context with alternative ways of exploiting low-S/N spectroscopy.

\subsection{What TSN is really learning}
\label{sec:discussion-mechanism}
TSN is trained to reproduce the specific high-S/N fluxes and ASPCAP label scale delivered by APOGEE for empirical spectral twins.
This paired objective teaches the encoder which spectral patterns are stable across visits versus those that vary stochastically.
To distinguish what is actually noise and avoid hallucinating features, it is crucial to utilise a large repository of low-S/N and high-S/N pairs from all sorts of sources (see Sec.~\ref{sec:data-pairs}).
This dependence can be quantified directly for the denoising pretraining stage: when we retrain the denoiser on random fractions of the flux-training set, the median reconstruction error drops sharply up to a training fraction of ${\sim}0.2$ and then largely saturates by ${\gtrsim}0.7$ (Appendix~\ref{sec:test-flux}, Fig.~\ref{fig:train_fraction}).
This learning curve indicates that TSN benefits from broad coverage of low-S/N visits, but that the adopted sample of paired spectra is already large enough that denoising performance is close to saturation.
However, stability is a necessary but not sufficient condition for astrophysical signal: systematic artefacts that persist across visits (e.g. detector persistence or stable telluric residuals) will be learned as part of the target model, meaning TSN suppresses stochastic error but may entrench systematic error.

\modified{A further caveat concerns the interpretation of the reconstructed flux itself.
The high-S/N combined spectrum is a practical empirical reference, but it is not a literal single-epoch ground truth for any individual visit.
In addition to instrumental and reduction effects, some stars---including a subset of red giants---can exhibit genuine epoch-to-epoch spectral variability associated with magnetic activity, chromospheric changes, or rotation.
Such variability is likely to set an irreducible floor on the achievable visit-to-combined reconstruction accuracy for some stars, even for an otherwise well-trained model.
Quantifying the corresponding contribution to the flux residuals is beyond the scope of the present work, but it should be kept in mind when interpreting TSN reconstructions for individual objects.}

At the label level, the estimator implements a form of statistical regularisation: TSN trades increased bias for reduced variance.
This regularisation manifests as a regression to the mean, effectively suppressing noise-driven fluctuations but potentially over-smoothing genuine spectral peculiarities in rare objects, especially for stars whose true labels lie far outside the training distribution.
As shown by \citet{2025OJAp....8E..95T}, this compression of dynamic range is a fundamental property of machine learning models when the input features (here, the noisy spectra) have significant measurement uncertainties.
This systematic underestimation of extreme values cannot be eliminated simply by increasing the training sample size.
The science tests in Section~\ref{sec:result} demonstrate that this trade-off preserves population-level structure, recovering distinct chemical sequences in the inner-disk/bulge and satellite systems rather than forcing them onto a single global manifold.

\subsection{Extending APOGEE's chemical reach}
\label{sec:discussion-gains}
The primary scientific payoff of TwinSpecNet is data recovery: by suppressing noise-driven label scatter in the $\mathrm{S/N_{visit}}\sim 20$--60 regime, TSN expands the set of APOGEE observations that can be used as chemically informative constraints rather than discarded or treated as upper limits.
In APOGEE DR17, this corresponds to of order $10^5$ stars in the low-S/N tail where standard analyses are often limited by statistical noise (Fig.~\ref{fig:snr}).
These gains manifest differently across science applications: as tighter within-cluster abundance dispersions without compressing between-cluster offsets (Fig.~\ref{fig:cluster}), as cleaner separation of inner-disk and bulge chemical sequences and suppression of the Al finger artefact (Fig.~\ref{fig:bulge}), as improved consistency with external literature and chemical-evolution models for satellite systems (Fig.~\ref{fig:subpop}), and as improved C/N-based age precision that approaches high-S/N performance (Fig.~\ref{fig:cn_age}).
Critically, TSN does not erase population-level structure: system-specific abundance offsets, satellite turnover features, and C/N correlations are preserved, indicating that paired learning regularizes noise without forcing stars onto a global manifold.

TSN's improved low-S/N performance opens several avenues for future work.
First, the tightened C/N precision suggests that TSN-derived abundances could enable more robust age estimates for the full APOGEE red giant sample.
Empirical C/N--age calibrations \citep[e.g.][]{2023A&A...678A.158A} typically require high-quality abundances to mitigate noise-driven scatter, limiting their applicability to the brightest targets.
By extending reliable C and N measurements into the low-S/N regime, TSN could support age-resolved chemo-dynamical studies across a larger fraction of the Galactic disc and halo, improving constraints on radial migration, disc heating, and halo formation timescales.
Second, the sharper $\alpha$-element sequences recovered in satellite systems can improve constraints on their star formation histories.
Chemical-evolution models predict that the location and sharpness of the different abundance distribution encode the star formation efficiency and gas outflow timescales \citep{2021ApJ...923..172H}, but observational constraints are often limited by abundance scatter in faint members.
TSN's noise suppression enables finer detection of turnover features and tighter comparisons to flexCE and similar models, potentially resolving degeneracies between star formation history parameters and clarifying the role of environmental quenching in the Magellanic Clouds and other satellites.

\subsection{Extending paired learning to other surveys}
\label{sec:discussion-compare}
Taken together, the tests above suggest that TSN's paired-learning recipe is not unique to APOGEE but illustrates a more general way of extracting information from low-S/N spectroscopy, complementary to emerging transformer-based foundation models for stellar spectra \citep[e.g.][]{2024MNRAS.527.1494L}.
Whenever a survey delivers repeated visits, deep calibration fields, or overlap regions with higher-S/N spectra, those observations define empirical low-/high-S/N pairs that can supervise denoisers and label estimators without synthetic noise injection.
The main requirements are a stable reference label scale for the high-S/N spectra, sufficient coverage of the parameter space of interest, and a training set that samples the range of instrumental systematics and observing conditions encountered in science fields.
Under those conditions, paired learning turns survey redundancy into signal and encourages observing strategies that deliberately place repeat observations where faint targets are most scientifically valuable.

These ingredients are already present or planned in several large spectroscopic programmes: ongoing APOGEE/SDSS-V observations include repeated fields and deep calibration plates, and optical surveys such as WEAVE and 4MOST are designed with dedicated calibration and overlap regions \citep{2017arXiv171103234K,2014SPIE.9147E..0LD,2019Msngr.175....3D}.
\modified{Gaia is also expected to expand its spectroscopic content in future releases, potentially providing more time-resolved spectral information that could be amenable to related paired or multi-epoch analyses \citep{2023A&A...674A...1G}.}
In such settings, a TSN-like model could be trained once on carefully selected pairs and then applied to single-visit spectra for the wider survey population, extending chemically informative samples to fainter stars while remaining anchored to the survey's internal abundance scale.

\section{Conclusion}
\label{sec:summary}
We introduced TwinSpecNet (TSN), an empirical paired-learning framework that couples Vision Transformer encoders with sequential flux reconstruction and label prediction objectives.
Trained on $3.2$ million visit--combined flux pairs ($141{,}003$ unique stars) and $550{,}354$ labeled visits, TSN learns survey-specific noise patterns while remaining anchored to APOGEE's ASPCAP scale.
The sequential training strategy---pretraining on flux reconstruction before fine-tuning for label inference---allows TSN to build robust spectral representations from the full diversity of paired observations before specializing to the labeled set.

On held-out data, TSN reduces label scatter at $\mathrm{S/N_{visit}}\sim 20$--60 with calibrated uncertainties that broadly track empirical dispersion.
Internal consistency tests show TSN tightens within-cluster abundance dispersions by $\sim$40\% while preserving between-cluster variance to within $\sim$5\%, indicating that the improvement is not driven by compression of abundance space.
External validation against SAGA high-resolution literature abundances confirms TSN follows the one-to-one relation and reduces scatter for several $\alpha$ elements at the metal-poor edge ($\feh \gtrsim \unit[-2.3]{dex}$).
Demonstrations across inner-disk and bulge populations, satellite systems (LMC, SMC, Sagittarius), and C/N-based age-dating show that TSN recovers cleaner chemical sequences, suppresses known artefacts (e.g.\ the Al finger), and improves age precision from $\sim\unit[1.70]{Gyr}$ to $\sim\unit[1.49]{Gyr}$ without erasing population-level abundance structure.

By demonstrating that empirical paired learning can extend chemical reach into APOGEE's low-S/N regime, TSN provides a template for exploiting repeated observations in other spectroscopic surveys (WEAVE, 4MOST, SDSS-V).
This approach turns survey redundancy into signal, suggesting observing strategies that prioritize repeat visits of scientifically valuable faint targets.
The primary limitation---regression toward the training-set mean for extreme objects---is intrinsic to supervised learning with noisy inputs, but population-level tests confirm TSN preserves distinct chemical enrichment histories across Galactic components.

\section*{Data availability}

This work uses data from the Sloan Digital Sky Survey IV (SDSS-IV) APOGEE-2 Data Release 17 (DR17) \citep{2023ApJS..267...44A}, available at \url{https://www.sdss.org/dr17/}.

\begin{acknowledgements}

\newline
{\it Facility:} APOGEE
\end{acknowledgements}

\begin{appendix}
\section{Flux reconstruction on held-out spectral twins}
\label{sec:test-flux}

\begin{figure*}[ht!]
\centering
\includegraphics{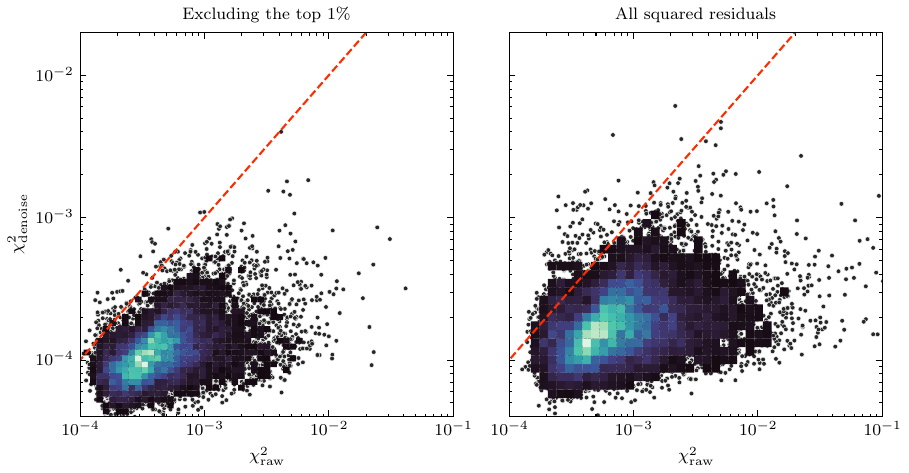}
\caption{
\modified{Denoising performance on held-out spectral twins.
Each point represents one held-out visit--star pair; the $x$-axis shows $\chi^2_\mathrm{raw}$, the mean squared residual between the raw low-S/N visit and its high-S/N twin, and the $y$-axis shows $\chi^2_\mathrm{denoise}$, the mean squared residual between the TSN reconstruction and the same reference.
Points below the red dashed one-to-one line indicate that TSN brings the spectrum closer to the high-S/N reference than the raw visit.
The left panel excludes the top 1\% of squared residuals per spectrum to reduce the impact of rare catastrophic pixels (e.g. cosmic rays or imperfectly masked bad pixels), while the right panel shows the same comparison without any clipping.
The same overall trend is present in both panels, demonstrating that the conclusion that TSN improves the majority of held-out spectra does not depend on this exclusion.}
}
\label{fig:denoise}
\end{figure*}

\modified{On the held-out paired sample, the TSN denoising branch reproduces the detailed line structure of the high-S/N reference spectra across a wide range of visit S/Ns.
We quantify reconstruction performance by comparing $\chi^2_\mathrm{raw}$ and $\chi^2_\mathrm{denoise}$, the mean squared residuals of the raw visit and the TSN reconstruction relative to the same high-S/N reference spectrum, respectively.
Fig.~\ref{fig:denoise} plots these two quantities directly against each other; points below the one-to-one line indicate that TSN reduces the mismatch to the reference.
The left panel excludes the top 1\% of squared residuals per spectrum to reduce the impact of rare catastrophic pixels (e.g. cosmic rays or imperfectly masked bad pixels).
The right panel shows the same comparison without any clipping and therefore provides the more conservative view.
The same overall trend is present in both panels, demonstrating that the conclusion that TSN improves the majority of held-out spectra does not depend on this exclusion.}

\begin{figure}[ht!]
\centering
\includegraphics{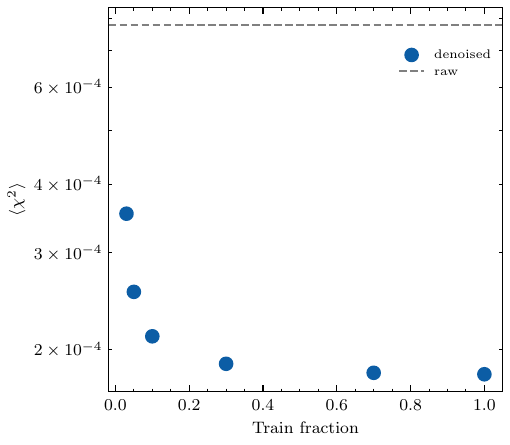}
\caption{
\modified{Dependence of denoising pretraining performance on the size of the paired flux-training set.
For each training fraction, we retrain the denoising branch on a random subset of the full flux-training sample and evaluate the median $\chi^2_\mathrm{denoise}$ on the same held-out visit--star test set, where each spectrum's $\chi^2$ is computed from all squared residuals relative to the high-S/N reference.
Lower values indicate better reconstructions.
The grey dashed line marks the corresponding median $\chi^2_\mathrm{raw}$ for the unprocessed visit spectra.
}
}
\label{fig:train_fraction}
\end{figure}

\modified{To test whether TSN depends critically on having extremely large numbers of paired spectra, we repeated the denoising pretraining using random fractions of the full flux-training sample and evaluated the resulting models on the same held-out paired test set.
Fig.~\ref{fig:train_fraction} shows the median $\chi^2_\mathrm{denoise}$ for each training fraction, computed from all squared residuals relative to the high-S/N reference, together with the corresponding raw-spectrum baseline.
This experiment is intended to probe the data dependence of the denoising pretraining stage, which provides the encoder initialization for the final label models, rather than to measure label accuracy directly as a function of training fraction.
The reconstruction error decreases steeply between training fractions of 0.03 and ${\sim}0.2$, showing that TSN benefits strongly from moving beyond the small-data regime and from seeing a broader diversity of paired visit spectra.
Beyond that point the curve flattens, and the additional improvement above ${\sim}0.7$ of the full sample is modest.
This behaviour implies that performance does depend on the breadth of the paired training set, but also that the adopted flux-training sample is already sufficient to capture most of the achievable gain for the denoising objective.
}

\section{Sensitivity to physical features}
\label{sec:test-interpret}
To connect TSN predictions to physical H-band diagnostics, we compute gradient spectra that quantify how each label depends on the input flux as a function of wavelength, analogous to saliency maps and attribution methods used in deep learning \citep[e.g.][]{2017arXiv170301365S}.
For a given spectrum, we evaluate the per-label saliency $\partial \mu_i / \partial f_\lambda$ of the predicted label with respect to the continuum-normalised flux.

\begin{figure*}[ht!]
\centering
\includegraphics{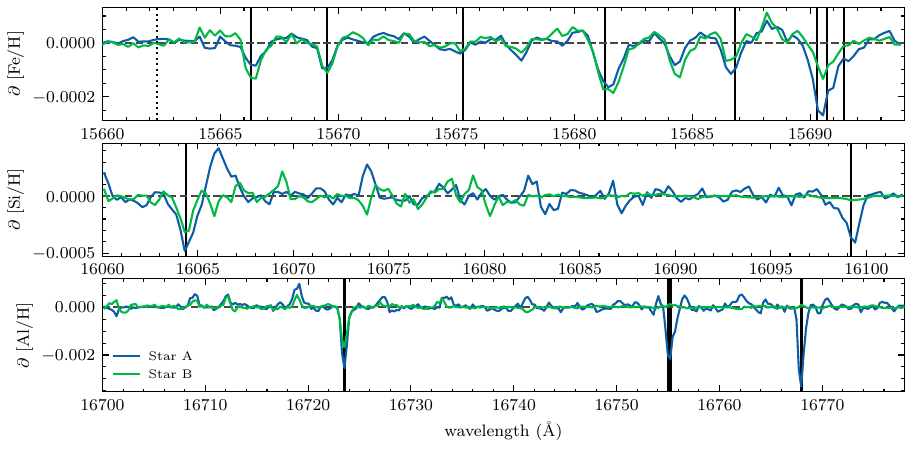}
\caption{
Partial derivatives of the three abundances (\feh, \xh{Si} and \xh{Al}) from TSN with respect to input wavelength for two representative stars with different stellar parameters.
Star A (blue): $T_\mathrm{eff}=\unit[4942]{K}$, $\log g=\unit[3.55]{dex}$, $\feh=\unit[-0.36]{dex}$; Star B (green): $T_\mathrm{eff}=\unit[3928]{K}$, $\log g=\unit[0.49]{dex}$, $\feh=\unit[-1.25]{dex}$.
Vertical lines show transitions from the APOGEE DR17 linelist \citep{2016AJ....151..144G,2021AJ....161..254S}: solid lines indicate ground-state transitions, while dotted lines mark Fe~II transitions.}
\label{fig:response}
\end{figure*}

Fig.~\ref{fig:response} shows representative gradient spectra for \feh, \xh{Si} and \xh{Al} for two giants with different stellar parameters.
Star A (blue, \texttt{APOGEE ID} = 2M15570684-1054133): $T_\mathrm{eff}=\unit[4942]{K}$, $\log g=\unit[3.55]{dex}$, $\feh=\unit[-0.36]{dex}$; Star B (green, \texttt{APOGEE ID} = 2M16544330+3932252): $T_\mathrm{eff}=\unit[3928]{K}$, $\log g=\unit[0.49]{dex}$, $\feh=\unit[-1.25]{dex}$.
We visualise these gradients alongside standard APOGEE \texttt{Turbospectrum} atomic and molecular line lists \citep{2016AJ....151..144G,2021AJ....161..254S}, which allows prominent sensitivity peaks to be associated with known atomic and molecular transitions.

TSN's label predictions respond most strongly at wavelengths corresponding to known atomic transitions of the respective elements, confirming that the model has learned to extract abundance information from physically motivated spectral diagnostics rather than from spurious correlations or continuum features.
The gradient amplitudes and patterns vary systematically with stellar parameters, and not all atomic lines contribute equally to the inferred abundances.
For example, the metal-poor giant (Star B, green) shows different sensitivities compared to the more metal-rich giant (Star A, blue), reflecting the physical reality that line strengths and blending depend on atmospheric conditions.
Similarly, some Al~I lines near $\unit[16755]{\AA}$ show weak or negligible gradients.
These patterns indicate that TSN is not merely memorizing training labels but has learned representations tied to interpretable H-band diagnostics, providing a safeguard against over-reliance on non-physical features and distinguishing TSN from fully black-box models.

\end{appendix}

\end{document}